
\magnification=1200
\baselineskip=18pt
\rightline{LPTB 93-1}
\centerline{Anyons as Dirac Strings, the $A_x=0$ Gauge}
\vskip 12pt
\centerline{John McCabe}
\centerline{Laboratoire de Physique Th\'eorique,\footnote{$^1$}{Unit\'e
de Recherche associ\'ee au CNRS n$^o$ 764.}
Universit\'e Bordeaux I}
\centerline{19 rue du Solarium, 33175 Gradignan FRANCE}
\vskip 20pt
\centerline{\bf ABSTRACT}
\vskip 15pt
\par We show how to quantize the anyon particle theory in a gauge,
$A_x=0$, where the statistical potential $\vec A(\vec x)$ is a
Dirac string. In this gauge, anyons obey normal
statistics.

\vfill\eject

\par Attempts to study the non-relativistic anyon model [1] have centered
on 2 methods. Either anyons are taken
to be normal bosons (or fermions) interacting with
a statistical potential in the coulomb gauge
($\partial^aA_a(\vec x)=0$), or anyons are taken to be free particles
obeying exotic statistics, a singular gauge [1,2].
There exists a third possibility which may
be more useful especially in systems having discrete translation
symmetries (ex. anyonic crystals). It involves taking anyons
to be bosons (or fermions) interacting
with a statistical potential having
the form of a Dirac string [3]. We will study
this possiblity which corresponds to the gauge $A_x(\vec x)=0$.
\par An anyon is a charged flux line in 2+1 dimensions. We start in
constructing the vector potential for a unit flux line located at the
origin. The associated vector potential satisfies the
equation $\vec\partial\times\vec A(\vec x)=\delta^2(x)$.
The potential
$(A_x(\vec x), A_y(\vec x))=(0,
{1\over 2}\delta(y)\epsilon(x))$,
$\epsilon(x)=\pm 1$ for $x{>\atop<}0$,
satisfies the gauge condition $A_x(\vec x)=0$ and
gives the correct magnetic field.
This potential has Dirac string
singularities along the positive and negative
$x$-axes and vanishes elsewhere.
Thus, the $A_x(\vec x)=0$ gauge corresponds to treating
anyons as pure Dirac strings.
We can write the Hamiltonian for N-anyons by taking a combination
of such Dirac strings. For N free non-relativistic
anyons, it is given by:
$$H_{anyon}={1\over 2m}\sum^N_{i=1}\left[-\partial^2_{x_i}+
(-i\partial_{y_i}+
{\alpha\over 2}\sum_{j\ne i}\delta(y_i-y_j)\epsilon(x_i-x_j)
)^2\right]\eqno{(1)}$$
\par Our gauge choice is not a singular gauge in the sense that the
N-particle wavefunction of (1) satisfies normal statistics. To
see this more clearly,
we will relate the $A_x=0$ gauge to the singular
gauge\footnote{$^2$}{In our conventions,
$\alpha=0$ corresponds to bosons
and $\alpha=2\pi$ to fermions.} ($\vec A(\vec x)=0$)
and to the coulomb gauge
($\partial^aA_a=0$) where the potential has the form $\vec A(r,\theta)=
{1\over 2\pi}\vec\partial\theta$ [4]. We start by showing
that the statistical potential in the $A_x=0$ gauge
can also be written as the
gradient of a "singular" gauge transformation. The transformation is
singular in the sense that the gauge
parameter is not a single-valued function on the plane, i.e. periodic in
$\theta$. It is a well-defined function on $\theta$'s covering space,
i.e. $\theta\epsilon{\bf R}$.
This was exactly the situation for the anyon
model in the Coulomb gauge where
$\vec A(\vec x)=\vec\partial({1\over 2\pi}\theta)$.
The parameter $\theta$ is a function on the covering space.
\par
 The potential, $A_x(\vec x)=0$ and $A_y(\vec x)={1\over 2}\delta(y)
\epsilon(x)$ can be written
as a gradient of a staircase function $\Omega(\theta)$.
$$\vec A(\vec x)=\vec\partial\Omega\ ,\ \ \ \ \ \
\Omega(\theta)\equiv {+m\over 2}\
\ \ \ {\rm for} \ \ \  \theta\epsilon[2m\pi,(2m+1)\pi), \
\ \ m\epsilon{\bf Z} \eqno{(2)}$$
To see this, write the gradient in angular coordinates,
$(\partial_x,\partial_y)=$
$(\cos\theta\partial_r-\sin\theta\partial_\theta,
\sin\theta\partial_r+\cos\theta\partial_\theta)$. $\Omega(\theta)$ is a
multi-valued function on the plane.
We can make a singular transformation, $\vec A^\prime=\vec A-
\vec\partial\Omega\equiv 0$, to
pass from the formulation of anyons as normal particles carrying Dirac
strings to the formulation as free particles obeying exotic statistics
($\alpha$ will determine the statistics).
We can also combine
two gauge transformations, $\vec A^\prime(\vec x)=
\vec A(\vec x)+\vec\partial\omega(\theta)$, with $\omega(\theta)\equiv
({1\over 2\pi}\theta-\Omega(\theta))$, to
transform from the $A_x=0$ gauge to the
$\partial^aA_a$ gauge. This does not effect the
the statistics, because the parameter $\omega(\theta)$ satisfies
$\omega(\theta\pm\pi)=\omega(\theta)$.
Thus, the gauge of the Hamiltonian in (1)
is not a singular gauge [1,2].
In what follows, we will
suppose that the N-particle wavefunction is bosonic.
\par We will now illustrate the $A_x=0$ gauge formulation
by finding the spectrum of 2 anyons confined to a circular box [5].
Since the Hamiltonian of (1) is
free for $x\ne0$ ($\vec x\equiv{1\over\sqrt 2}(\vec x_1-\vec x_2)$),
we can write the eigenfunction for the relative problem as:
$$\eqalign{\Psi_{\gamma k}
(r,\theta)&=\exp(i\gamma\theta)J_{|\gamma|}(kr)
\ \ \ \ {\rm for}\ y>0, \cr
\Psi_{\gamma k}
(r,\theta)&=A\exp(i\beta\theta)J_{|\beta|}(kr)
\ \ \ \ {\rm for} \ y<0.\cr
}\eqno{(3)}$$
The energy, $E={k^2\over 2m}$, is fixed by the boundary
condition on the disc, $\Psi_{\gamma k}(r=R,\theta)=0$, which implies
that $J_{|\gamma|}(kR)=0$.
\par To determine A
and $\beta$, we must impose that the wavefunction is bosonic.
Our choice of the statistical potential preserves the invariance of
$H_{anyon}$ under the exchange of two coodinates $\vec x_i$ and
$\vec x_j$. Thus, Bose statistics is implemented by requiring that
$\Psi(\vec x)=\Psi(-\vec x)$ or in angular coordinates that
$\Psi(r, \theta)=\Psi(r,\theta+\pi)$ with
$\theta\epsilon [0,2\pi)$. This condition determines the wavefunction
for $y<0$ in terms of the wavefunction for $y>0$. The solutions are:
$$\eqalign{\Psi_{\gamma k}
(r,\theta)&=\exp(i\gamma\theta)J_{|\gamma|}(kr)
\ \ \ \ {\rm for}\ y>0, \cr
\Psi_{\gamma k}
(r,\theta)&=\exp(i\gamma\theta-i\tilde\gamma\pi)J_{|\gamma|}(kr)
\ \ \ \ {\rm for} \ y<0.\cr
}\eqno{(4)}$$
Where, $\tilde\gamma=\gamma -2m$
with $m$ any integer. The wavefunctions
(4) satisfy all the constraints of statistics. The spectrum is
determined by the quantum number $\gamma$. $\gamma$
is fixed by the equation of motion on the $x$-axis.
To see how (4) solves the Hamiltonian equations on the $x$-axis, we
remark that the wavefunction has phase discontinuities, $e^{-i\tilde
\gamma\pi}$, in crossing the either the
negative or positive $x$-axis in a clockwise sense.
For example, in a neighborhood of the negative $x$-axis,
we can write (4) as:
$$\exp{\left[i{\tilde\gamma\pi\over 2}(\epsilon(y)-1)\right]}e^{i\gamma
 \theta}J_{|\gamma |}(kr)\eqno{(5)}$$
Substituting this form into (1), we find:
$$k^2e^{i\gamma\theta}J_{|\gamma|}(kr)=
\left[-\partial^2_x +(-i\partial_y +[\tilde\gamma\pi+{\alpha\over 2}]
\delta(y))^2\right]e^{i\gamma\theta}J_{|\gamma|}(kr)\eqno{(6)}$$
This equation only has a solution if the coefficient of the $\delta(y)$
vanishes.  Thus, we arrive at a condition on $\tilde\gamma$,

$\tilde\gamma={\alpha\over 2\pi}$ or
$\gamma={\alpha\over 2\pi}+2m$. The equatoin of motion on the $x$-axis
has determine the non-integer part of $\gamma$. This shows
that the 2-anyon relative eigenfunctions, in the $A_x(\vec x)=0$
gauge, are given by:
$$\eqalign{\Psi_{m k}
(r,\theta)&=\exp(i[{\alpha\over2\pi}+2m]\theta)
J_{|{\alpha\over 2\pi}+2m|}(kr)
\ \ \ {\rm for}\ \ \ y>0, \cr
\Psi_{m k}
(r,\theta)&=\exp(i[{\alpha\over 2\pi}+2m]\theta-i{\alpha\over 2})
J_{|{\alpha\over 2\pi}+2m|}(kr)
\ \ \ {\rm for} \ \ \ y<0\cr}\eqno{(7)}$$
Here, the states have been labeled by the two traditional quantum numbers
$m\epsilon {\bf Z}$ and $k$.
We would arrive at the same conclusions in substituting (4) in (1) on a
neighborhood of the positive $x$-axis.
The values of $k$ and the
energy eigenvalues are given by the zeros of the Bessel function.
$$E_{m}(k)={1\over 2m}\left({z_{sm}\over R}\right)^2 \ \ \
k={z_{sm}\over r}\ \ \
{\rm with}\ J_{|{\alpha\over 2\pi}+2m|}(z_{sm})\equiv 0\ \ \
s=1,2,3\ldots \eqno{(8)}$$
This agrees with the results of [5] in the coulomb and singular gauges.
\par The $A_x(\vec x)=0$ gauge has free particle wavefunctions with
phase discontinuities, $e^{- i {\alpha\over 2}}$ when any 2 anyons have
equal $y$-coordinates. This gauge may be more useful than the coulomb
gauge in problems having discrete
symmetries in the $x-$ and $y-$directions, ex.
crystal of anyons (or flux tubes)
where one may want to expand the anyonic interaction
in a unit cell by a Fourier series. These possibilities are yet to be
investigated.

\vskip 15pt
\centerline{\bf REFERENCES}
\vskip 15pt
\item{[1]}J.M. Leinaas and J. Myrheim, Nuovo Cim. B 37 (1977) 1;
J.M. Leinaas, Nuovo Cim. A 47 (1978) 1; M.G.G. Laidlaw and C.M. de Wit,
Phys. Rev. D 3 (1971) 1375.
\item{[2]}F. Wilczek, Phys. Rev. Lett. 48 (1982)
1144; ibid 49 (1982) 957.
\item{[3]}P.A.M. Dirac, Proc. of the Royal Soc. of London A133 (1931)
 60; ibid Phys. Rev. 74 (1948) 817; ibid Int. J. Theor. Phys. 17 (1978)
235.
\item{[4]}Y. Aharonov and D. Bohm, Phys. Rev. 115 (1959) 485.
\item{[5]}D.P. Arovas, R. Schrieffer, F. Wilczek, and A. Zee, Nucl.
Phys. B251 (1985) 117; J.S. Dowker, J. Phys. A 18 (1985) 3521.
\vfill\eject\end